\documentclass{article}
\usepackage[utf8]{inputenc}
\usepackage{times}
\usepackage{amsfonts}
\usepackage{amsmath,amssymb}
\usepackage{indentfirst}
\usepackage{wrapfig}
\usepackage{graphicx}
\usepackage{graphics}
\usepackage{caption}
\usepackage{subcaption}
\usepackage[toc,page]{appendix}
\usepackage{color}
\usepackage{cite}
\usepackage[inactive]{srcltx}
\usepackage[T1]{fontenc}
\usepackage{float}
\usepackage{hyperref}
\newcommand{\sech}{\textrm{sech}}
\begin{document}
	\date{}
\begin{center}
	{\Large\bf Generation of nonclassical states of light via truncation of mixed states}
\end{center}
\begin{center}
	{\normalsize E.P. Mattos and A. Vidiella-Barranco \footnote{vidiella@ifi.unicamp.br}}
\end{center}
\begin{center}
	{\normalsize{Gleb Wataghin Institute of Physics - University of Campinas}}\\
	{\normalsize{ 13083-859   Campinas,  SP,  Brazil}}\\
\end{center}
\begin{abstract}
A possible way of generating nonclassical states of light, especially non-Gaussian states, is via the truncation of 
a given state in the Fock basis. In recent work, we presented an alternative scheme for such quantum scissors 
[Phys. Rev. A, {\bf 104}, 033715 (2021)], employing a nondegenerate parametric amplifier, a beam splitter and
photodetectors. An advantage of this setup is that it does not require the generation of Fock states beforehand, 
as in previous proposals. Here we extend this treatment to mixed input states. We show the possibilities of generating 
truncated states with either a maximum Fock number $N$ or states having a minimum Fock number $N$. We discuss two specific 
examples of states to be truncated: i) the thermal state, and ii) the phase-diffused coherent state. In both cases, we 
show that the generated states can have significant sub-Poissonian statistics as well as non-Gaussian character. 
The degree of such nonclassical properties, as well as the success probabilities, can be changed by adjusting the 
parametric amplifier strength and the beam splitter transmittance.
\end{abstract}
%
\section{Introduction}
\label{section1}
The generation of quantum states of light \cite{dodonov02} is of central importance for the development of quantum technologies 
\cite{browne17,barnett17}. States such as single-photon number (Fock) states or entangled states are essential resources 
for protocols of quantum communication, for instance. In the past years several quantum state engineering protocols have been 
proposed. Some of those schemes involve the interaction of light in special media \cite{kilin95,avb98a}, 
or in cavity QED systems \cite{schleich93,dantsker94,eberly96,avb98b}. We may also cite processes of photon addition \cite{agarwal91}, 
photon subtraction \cite{agarwal92}, the removal of specific components in the Fock basis ("hole burning") 
\cite{baseia04}, and the use of arrays of beam splitters followed by photodetections \cite{welsch99}. 
A particularly interesting operation is known as quantum state truncation, 
or "quantum scissors" \cite{barnett98}. In this process, a quantum state of light, $|\varphi\rangle$, is transformed in a state having a 
finite number of Fock components, the simplest one being a superposition of the vacuum state $|0\rangle$ and the one
photon state $|1\rangle$: $\hat{T}|\varphi\rangle = c_0|0\rangle + c_1|1\rangle$. The original quantum scissors device 
\cite{barnett98} requires the injection of a single-photon (Fock) state, the vacuum state, and an arbitrary field 
state $|\varphi\rangle$ in an arrangement containing two beam splitters. Photodetections are performed in two of the three output ports 
of the system, which allows the generation of a truncated state in the remaining port. Other types of truncated states can be generated 
in alternative configurations for the device, in which the injection of more than one Fock state is required \cite{imoto07,yuan20}.
The quantum scissors have been experimentally realized in the context of quantum teleportation \cite{lvovsky03}, and for the implementation 
of an optical noiseless amplifier \cite{grangier10}. More recently, we can find in the literature a growing number of works about possible 
applications of quantum scissors, for example: truncation of thermal states \cite{yuan17}, improvement of entanglement \cite{liao18,zubairy19}, 
continuous variable quantum key distribution \cite{razavi20}, quantum repeaters \cite{guha20}, and noiseless amplification \cite{ralph20,malaney21}.
It has also been shown that the implementation of quantum state truncation can be accomplished without nonclassical input states if one of 
the beam splitters is replaced by a parametric amplifier, as shown in \cite{scissors21}. This could simplify the experimental setup, since 
in the previous propositions for quantum scissors, it is necessary to generate the (highly nonclassical) Fock states beforehand. 
In addition to that, most of the discussions involving quantum scissors also assume pure input states of light, which may not  
be easily available to be used as initial states. It would be therefore interesting to investigate a scenario where noisier states 
are used as inputs to the quantum scissors device.

In this work, we are going to be concerned with generating nonclassical states of light via the truncation of mixed input states. 
For that, we consider the recently introduced alternative quantum scissors \cite{scissors21} with input states which are: a
mixed field state diagonal in the Fock basis, vacuum states and a classical pumping field. We will discuss states generated from two 
possible input states, namely: i) thermal states (super-Poissonian) and ii) phase-diffused coherent states (Poissonian). We will 
show that the resulting states may have nonclassical features such as sub-Poissonian statistics and non-Gaussian character, both of 
them of relevance for the development of quantum technologies. Here we will be quantifying these properties using Mandel's 
$Q$ parameter \cite{mandel79} and the Hellinger distance \cite{luo20}, respectively.

This paper is organized as follows: In Section \ref{section2} we obtain the state produced by the action of the modified quantum scissors, 
from a mixed state which is diagonal in the Fock basis. In Section \ref{section3} we consider two specific mixed input states as examples and discuss 
in some detail the non-classical properties of the generated states. In Section \ref{section4} we present our conclusions.

\section{Truncation of mixed states}
\label{section2}

\begin{figure}[htbp]
	\centering
	\includegraphics{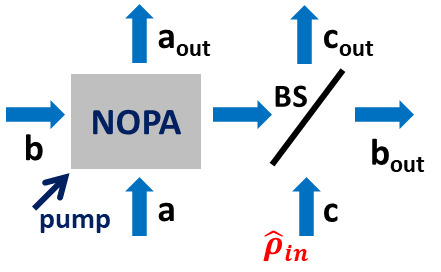}
	\caption{Schematic illustration of the proposed setup: a Nondegenerated Optical Parametric Amplifier (NOPA)
		has input modes $\hat{a}$ and $\hat{b}$ and a classical pump with strength $s$ and phase $\phi$.
		One of the output modes enters a beam-splitter (BS), which has a second input mode, $\hat{c}$. 
		Photodetectors may be placed in pairs in the output modes, either in $\hat{b}_{out}$ and $\hat{c}_{out}$ 
		or $\hat{a}_{out}$ and $\hat{c}_{out}$.}
	\label{fig:setupscissors}
\end{figure}

The alternative quantum scissors introduced in \cite{scissors21} allows quantum state truncation using only Gaussian input states. The 
proposed setup consists of a nondegenerate optical parametric amplifier and a beam splitter, as represented in Figure \ref{fig:setupscissors}. 
Here we assume that the input mode $\hat{c}$ is in a mixed state $\hat{\rho}_{in}$ diagonal in the Fock basis, while modes 
$\hat{a}$ and $\hat{b}$ are both in vacuum states, i.e., the joint initial state is
\begin{equation}
	\hat{\rho}_{in} = \sum_{n=0}^{\infty} \rho_n |0,0,n\rangle\langle 0,0,n|.
\end{equation}
The state generated by the parametric amplifier/ beam splitter will be
\begin{equation}
	\hat{\rho}_{out}=\hat{R}(\theta)\hat{S}(\xi)\hat{\rho}_{in}\hat{S}(\xi)^\dag \hat{R}(\theta)^\dag
	=\sum_{n=0}^\infty\rho_n \hat{R}(\theta)\hat{S}(\xi) |0,0,n\rangle\langle 0,0,n|\hat{S}(\xi)^\dag \hat{R}(\theta)^\dag.
\end{equation}
Here $\hat{S} = \exp\left[\xi^* \hat{a}\hat{b} - \xi \hat{a}^\dagger\hat{b}^\dagger \right]$ is the two-mode 
squeezing operator and $\hat{R} = \exp\left[i\theta(\hat{a}^\dagger\hat{b} + \hat{a}\hat{b}^\dagger)\right]$ 
the beam splitter operator. Also, $\xi = s e^{i\phi}$, where $s$ is the strength of the amplifier ($s \geq 0$), and $\phi$ 
the phase of the (classical) pump field. We have parametrized the transmittance $\mathtt{T}$ and reflectance 
$\mathtt{R}$ of the beam splitter $(|\mathtt{T}|^2 + |\mathtt{R}|^2 = 1)$ using the angle $\theta$, so that
$\mathtt{T} = \cos\theta$ and $\mathtt{R} = i\sin\theta$. After some algebra, we obtain the following tripartite field state
\begin{multline}
	\hat{\rho}_{out}  =\sum_{n=0}^\infty\sum_{k=0}^\infty\sum_{p=0}^\infty\sum_{l=0}^k\sum_{q=0}^p\sum_{m=0}^n
	\sum_{u=0}^n\frac{n!}{m!(n-m)!u!(m-u)!}\frac{\sqrt{k!}}{l!(k-l)!}\frac{\sqrt{p!}}{q!(p-q)!}\rho_n \\
	\times A_k(s,\phi)A_p^*(s,\phi)\sqrt{(k-l+n-m)!}\sqrt{(l+m)!}\sqrt{(p-q+n-u)!}\sqrt{(q+u)!} \\ 
	\times(-1)^{2n-m-u}\mathtt{T}^{m+p-q}\mathtt{T}^{*k-l+u}\mathtt{R}^{l+n-u}\mathtt{R}^{*n-m+q}|k,n-m+k-l,m+l\rangle\langle p,n-u+p-q,u+q|,
	\label{eq:statecomplete}
\end{multline}
with  $A_n(s,\phi)=\sech\,s(-e^{i\phi}\tanh{s})^n$. The interactions in the non-linear and linear optical elements are 
followed by photodetections in two different output ports, so that the generated state comes out from the remaining port. 
It is possible to generate different types of single-mode truncated states from the state in Equation (\ref{eq:statecomplete})
depending on where the photodetectors are placed, as we will show below.

\subsection{Photodetections at the \boldmath\texorpdfstring{$\hat{b}_{out}$}{} and \boldmath\texorpdfstring{$\hat{c}_{out}$}{} output ports}

Consider two photodetectors placed at the beam splitter output ports in such a way that $N$ photons are detected in $\hat{b}_{out}$ and
no photons are detected in $\hat{c}_{out}$. The resulting state will be
\begin{equation}
	\hat{\rho}_a^{(N,0)}=\frac{1}{p_a^{(N,0)}}\sech^2s\sum_{k=0}^N\frac{N!}{k!(N-k)!}\rho_{N-k}\tanh^{2k}s\ |\mathtt{T}|^{2k}|\mathtt{R}|^{2(N-k)}|k\rangle\langle k|,
	\label{eq:stateroa}
\end{equation}

with generation probability $p_a^{(N,0)}$

\begin{equation}
	p_a^{(N,0)}=\sech^2s\sum_{k=0}^N\frac{N!}{k!(N-k)!}\rho_{N-k}\tanh^{2k}s\ |\mathtt{T}|^{2k}|\mathtt{R}|^{2(N-k)}.
\end{equation}
The state in Equation (\ref{eq:stateroa}) is a mixed truncated state with a maximum Fock number $N$. 

\subsection{Photodetections at the \boldmath\texorpdfstring{$\hat{a}_{out}$}{} and \boldmath\texorpdfstring{$\hat{c}_{out}$}{} output ports}

We may also place one photodetector at output port $\hat{a}_{out}$ (parametric amplifier) and the other at output port $\hat{c}_{out}$ 
(beamsplitter). If $N$ and $0$ photons are detected, respectively, we obtain the following state,

\begin{equation}
	\hat{\rho}_b^{(N,0)}=\frac{1}{p_b^{(N,0)}}\sech^2s\ \tanh^{2N}s\ |\mathtt{T}|^{2N}
	\sum_{n=0}^\infty\frac{(n+N)!}{n!N!}\rho_n|\mathtt{R}|^{2n}|n+N\rangle\langle n+N|,
\end{equation}

with generation probability $p_b^{(N,0)}$

\begin{equation}
	p_b^{(N,0)}=\sech^2s\ \tanh^{2N}s\ |\mathtt{T}|^{2N}\sum_{n=0}^\infty\frac{(n+N)!}{n!N!}\rho_n|\mathtt{R}|^{2n}.
\end{equation}
In this case the generated state has a minimum Fock number $N$. We remark that it suffices to have $N = 1$, i.e., the 
removal of the vacuum state from a mixed state, to obtain a truncated nonclassical state \cite{lee95} $\hat{\rho}_b^{(1,0)}$. 

\section{Nonclassical properties of the generated states}
\label{section3}

Now, we would like to illustrate the proposed scheme for two different input (mixed) states, which are the 
thermal state $\hat{\rho}_{in}^{(th)}$, and the phase diffused coherent state $\hat{\rho}_{in}^{(pd)}$. We will focus our analysis 
on the properties of the truncated states exiting output port $\hat{a}_{out}$, that is, $\hat{\rho}_a^{(N,0)}$. 
Before starting our discussion, we would like to make some considerations about the states being generated. For simplicity, we take
the case of a single photodetection, i.e. $N = 1$,
\begin{equation}
	\hat{\rho}_a^{(1,0)} = \frac{1}{p_{a}^{(1,0)}}\sech^2 s
	\left(\rho_1 |\mathtt{R}|^2|0\rangle\langle 0| + \rho_0 |\mathtt{T}|^2\tanh^2{s}\ |1\rangle\langle 1|\right),\label{eq:truncated1}
\end{equation}
with
\begin{equation}
	p_{a}^{(1,0)} = \sech^2s\left(\rho_1 |\mathtt{R}|^2 + \rho_0 |\mathtt{T}|^2 \tanh^2s \right).
	\label{eq:probgena}
\end{equation}
Even though the state in Equation \ref{eq:truncated1} represents an incoherent superposition of the vacuum
state $|0\rangle\langle 0|$ with the one-photon state $|1\rangle\langle 1|$, it can have significant nonclassical features. 
We should remark that if $\mathtt{R} = 0\,\,(\theta = 0, \pi\, \ldots \mbox{rad})$, the generated state  
will be precisely the one-photon state. However the corresponding probability of generation happens to be lower 
than in other cases. Also, if $\mathtt{T} = 0,\,(\theta = \pi/2, 3\pi/2\, \ldots \mbox{rad})$, the generated state  
will be the vacuum state. We will disregard these extreme situations and rather concentrate our discussion 
on the truncated states having higher success probabilities.

\subsection{Thermal state input}

The single mode thermal state, the maximally mixed state having a fixed energy is also a Gaussian state, and has the following 
coefficients in the Fock basis
\begin{equation}
	\rho_n^{(th)} = \frac{1}{\Bar{n}_{th}+1}\left(\frac{\Bar{n}_{th}}{\Bar{n}_{th}+1}\right)^n.
\end{equation}
Here $\Bar{n}_{th} =  [\exp(\omega/k_B T)-1]^{-1}$ is the mean photon number of the state with mode frequency $\omega$ and 
effective temperature $T$. Note that in this case our quantum scissors device is being exclusively fed by Gaussian input states 
which are also "classical-like".

\begin{figure}[htbp]
	\centering
	\includegraphics[scale=0.50]{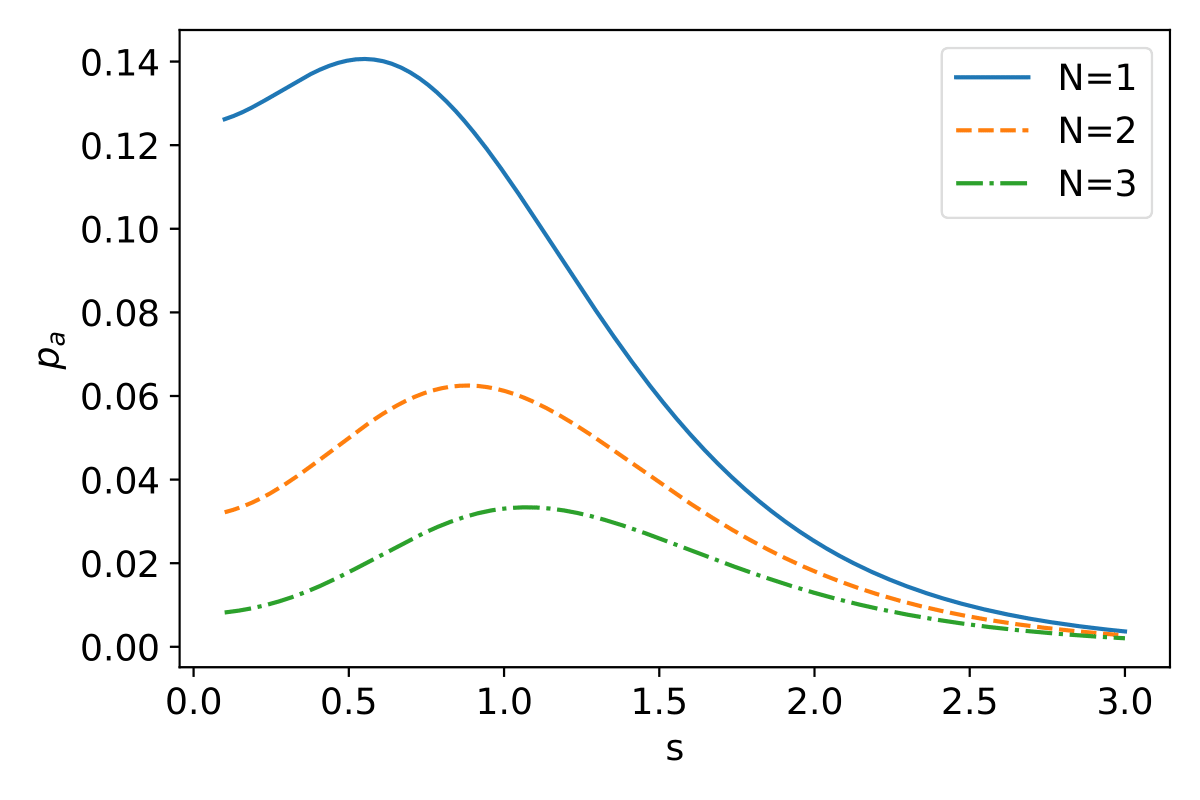}
	\caption{Generation probability $p_A^{(N,0)}$ of states $\hat{\rho}_a^{(N,0)}$ as a function of $s$ for an initial thermal 
		state with $\Bar{n}_{th} = 1.0$ and a $50 : 50$ beam splitter ($\theta = \pi/4$ rad).}
	\label{fig:prob_th_a}
\end{figure}

We would like to generate truncated states with interesting properties having as large as possible generation probabilities $p_a^{(N,0)}$.
This can be accomplished by adjusting the parameters of the scissors, e.g., the amplifier strength $s$. 
To illustrate this, we plotted in Figure \ref{fig:prob_th_a} the success probability $p_a$ as a function of $s$ for an input thermal 
state with $\Bar{n}_{th} = 1.0$ and a $50 : 50$ beam splitter ($\theta = \pi/4$ rad). This is done for $N = 1,\, 2,\, 3$. 
For larger values of $N$, as expected, we verify a significant decrease in $p_a$, also noting that $p_a^{(1,0)}$ is maximum for 
$s \approx 0.5$.

\subsubsection{Sub-Poissonian statistics: thermal state input}

\begin{figure}[htbp]
	\centering
	\includegraphics[scale=0.50]{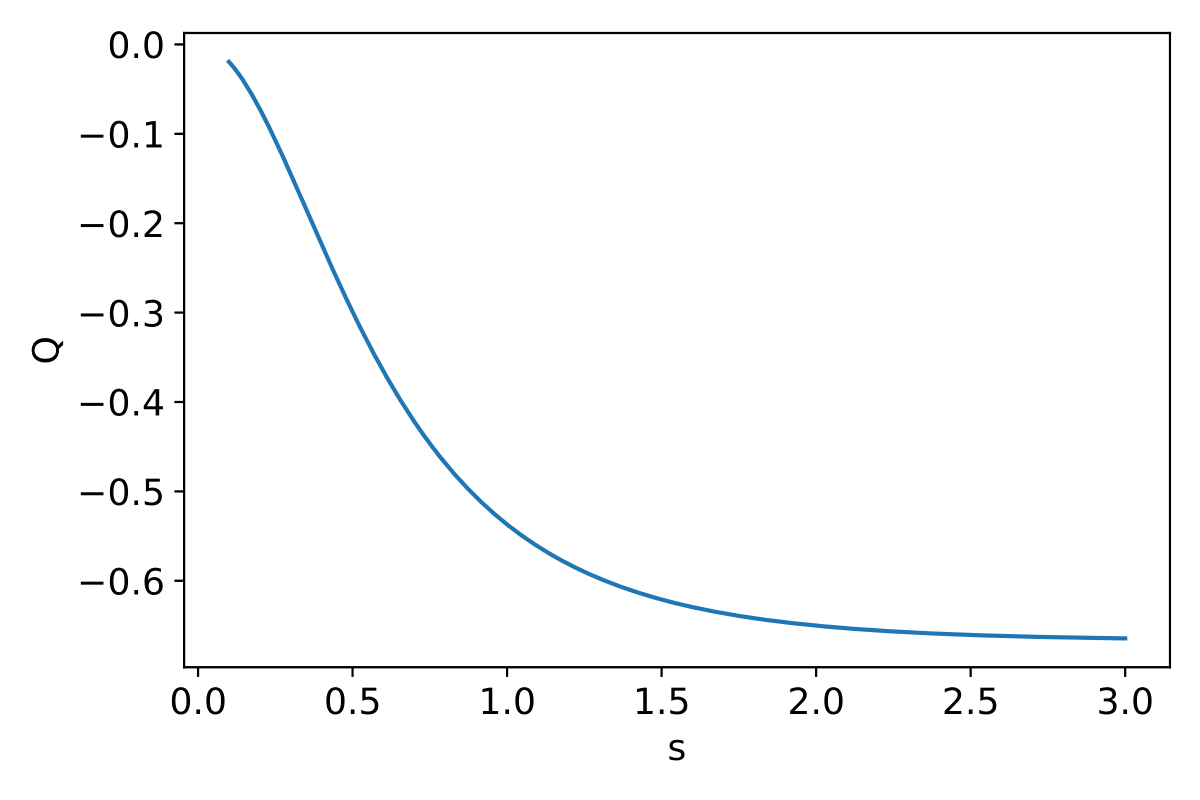}
	\caption{Mandel $Q$ parameter of state $\hat{\rho}_a^{(1,0)}$ as a function of $s$ for an initial thermal 
		state with $\Bar{n}_{th} = 1.0$ and a $50 : 50$ beam splitter ($\theta = \pi/4$ rad).}
	\label{fig:qparam_a_s_thermal}
\end{figure}

\begin{figure}[htbp]
	\centering
	\includegraphics[scale=0.50]{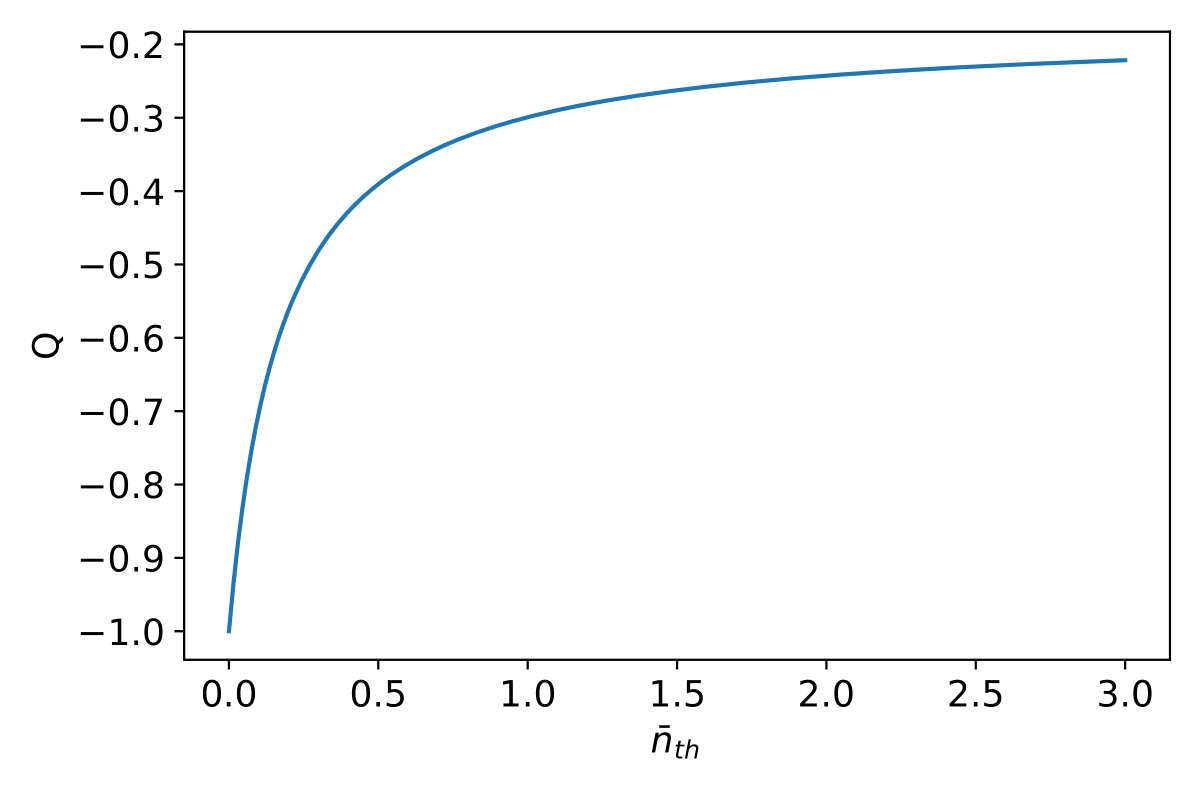}
	\caption{Mandel $Q$ parameter of state $\hat{\rho}_a^{(1,0)}$ as a function of 
		$\Bar{n}_{th}$ for an initial thermal state, with $s = 0.5$ and a $50 : 50$ beam splitter ($\theta = \pi/4$ rad).}
	\label{fig:qparam_a_n_thermal}
\end{figure}

A characteristic nonclassical feature of light fields is the sub-Poissonian character. It may be quantified by the Mandel $Q$ 
parameter, defined as $Q =  \left\langle (\Delta \hat{n})^2\right\rangle/\left\langle \hat{n} \right\rangle - 1$ \cite{mandel79}
and that naturally indicates deviations from the Poissonian photon statistics of a coherent state. It is null for coherent states 
and assumes the minimum value $Q = -1$ for states having an exact number of photons (Fock states). On the other hand,
thermal states have larger fluctuations in photon number, being for that reason called super-Poissonian states. 
For a thermal state with mean photon number $\Bar{n}_{th}$, Mandel's parameter is simply $Q = \Bar{n}_{th}$. We may now calculate the 
expectation values $\left\langle (\Delta \hat{n})^2\right\rangle \equiv Tr[\hat{\rho}_a^{(N,0)} (\hat{a}^\dagger\hat{a})^2] 
- (Tr[\hat{\rho}_a^{(N,0)} \hat{a}^\dagger\hat{a}])^2$ and 
$\left\langle \hat{n} \right\rangle \equiv Tr[\hat{\rho}_a^{(N,0)} \hat{a}^\dagger\hat{a}]$ and obtain the $Q$ parameter
for the state $\hat{\rho}_a^{(N,0)}$,
\begin{equation}
	Q = - \frac{\tanh^{2}s\ |\mathtt{T}|^2}{\frac{\Bar{n}_{th}}{1 + \Bar{n}_{th}}|\mathtt{R}|^2 + \tanh^{2}s\ |\mathtt{T}|^2}.
	\label{eq:qmandelrhoa}
\end{equation}
Notably, the Mandel $Q$ parameter of such truncated states does not depend on $N$. This peculiarity must be unique to the
thermal state and is certainly related to the specific form of its photon number distribution. 
In fact, for the initial state that will be discussed in the next section, Mandel's $Q$ parameter has a 
dependency on $N$. In Equation \ref{eq:qmandelrhoa}, we see that the states $\hat{\rho}_a^{(N,0)}$ are generally sub-Poissonian.
In Figure \ref{fig:qparam_a_s_thermal} we have plotted Mandel's $Q$ parameter relative to the state $\hat{\rho}_a^{(N,0)}$
as a function of $s$ for a $50:50$ beam splitter ($\theta = \pi/4$ rad) and $\Bar{n}_{th} = 1.0$. In our modified quantum scissors, 
we could generate a state with specific values of the $Q$ parameter by controlling both the amplifier strength $s$ and the beam 
splitter transmittance $\mathtt{T} = \cos\theta$, for instance. In this case, the larger the amplifier
strength $s$, the more sub-Poissonian becomes the generated field, although the success probability decreases, 
as seen in Figure \ref{fig:prob_th_a}. We may therefore optimize the device's parameters in order to obtain the desired 
results. The dependence of Mandel's $Q$ parameter on the mean photon number of the input thermal state, $\Bar{n}_{th}$, is shown in 
Figure \ref{fig:qparam_a_n_thermal}. As expected, for increasingly super-Poissonian initial states ($\Bar{n}_{th}$ increasing), 
the generated truncated state becomes less sub-Poissonian.

\subsubsection{Non-Gaussian character: thermal state input}

Non-Gaussian states are essential resources for the implementation of quantum technologies \cite{walschaers21}, and 
therefore it is of interest to assess the non-Gaussian character of the generated states. For that, we use a recently
introduced non-Gaussianity quantifier, based on the Hellinger distance \cite{luo20}. For a state having a density
operator $\hat{\rho}$, the non-Gaussianity $H$ is defined as
\begin{equation}
	H = \sqrt{1-\mbox{Tr}\sqrt{\hat{\rho}}\sqrt{\hat{\rho}_g}},
\end{equation}
where $\hat{\rho}_g$ is the Gaussian reference state, that is, the unique state that has the same average and
covariance matrix as $\hat{\rho}$ does. The quantifier $H$ is such that $0 \leq H \leq 1$, being null if the state 
$\hat{\rho}$ is a Gaussian state.

\begin{figure}[htbp]
	\centering
	\includegraphics[scale=0.50]{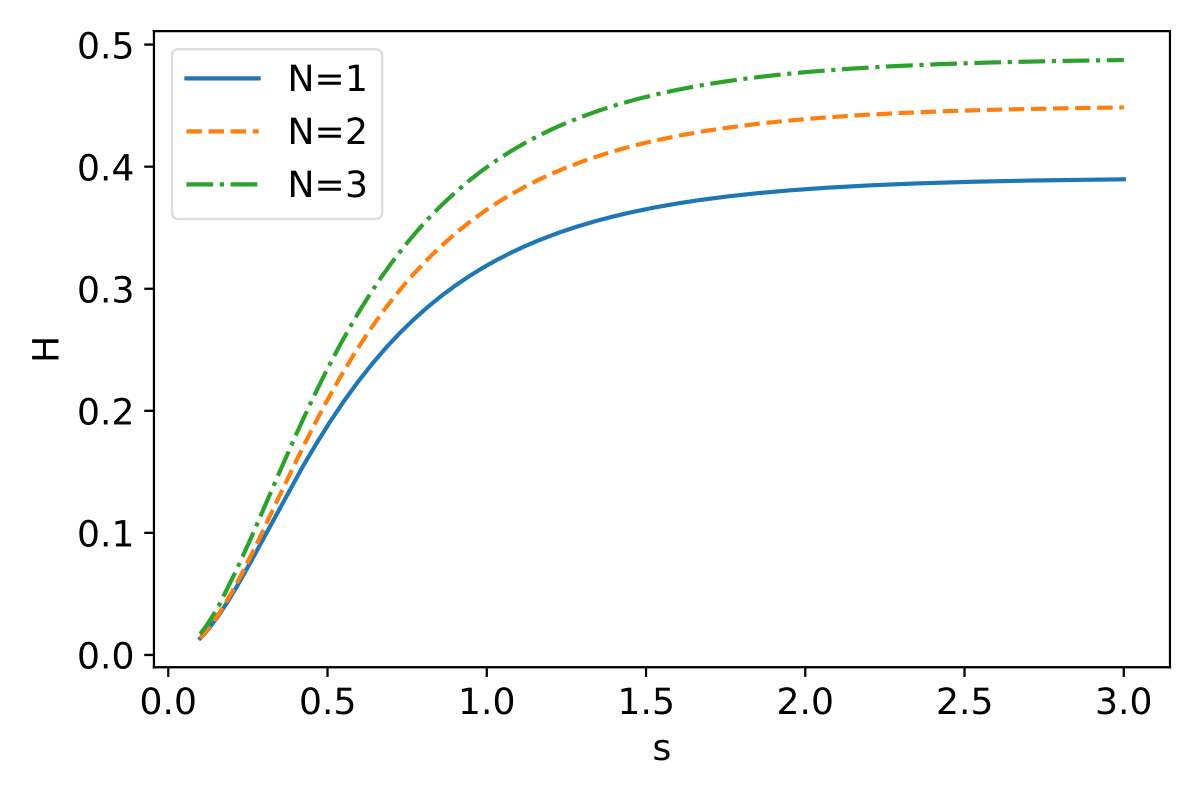}
	\caption{Non-Gaussianity quantifier $H$ of states $\hat{\rho}_a^{(N,0)}$ as a function of $s$ for an initial thermal 
		state with $\Bar{n}_{th} = 1.0$ and a $50 : 50$ beam splitter ($\theta = \pi/4$ rad).}
	\label{fig:H_a_s_thermal}
\end{figure}

\begin{figure}[htbp]
	\centering
	\includegraphics[scale=0.50]{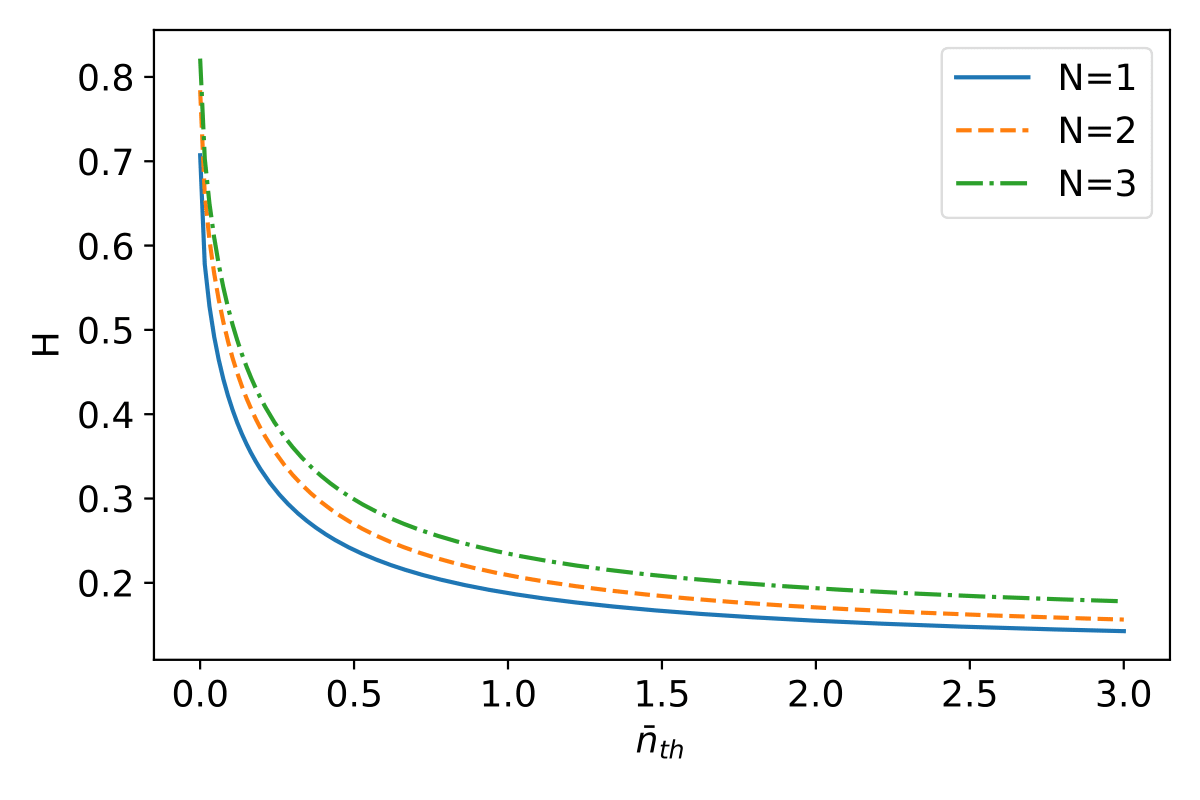}
	\caption{Non-Gaussianity quantifier $H$ of states $\hat{\rho}_a^{(N,0)}$ as a function of 
		$\Bar{n}_{th}$ for an initial thermal state, with $s = 0.5$ and a $50 : 50$ beam splitter ($\theta = \pi/4$ rad).}
	\label{fig:H_a_n_thermal}
\end{figure}

For field states diagonal in the Fock basis, $\hat{\rho} = \sum_{n=0}^\infty p_n|n\rangle\langle n|$, $H$
reads
\begin{equation}
	H = \sqrt{1-\sum_{n=0}^\infty\sqrt{\frac{p_n\Bar{n}^n}{(\Bar{n}+1)^{n+1}}}},
\end{equation}
with $\Bar{n}$ being the mean photon number in state $\hat{\rho}$. In Figure \ref{fig:H_a_s_thermal} 
we have plotted the non-Gaussianity quantifier $H$ of state $\hat{\rho}_a^{(N,0)}$
as a function of $s$ for a $50:50$ beam splitter ($\theta = \pi/4$ rad) and $\Bar{n}_{th} = 1.0$. We note that
the generated state becomes more non-Gaussian as the amplifier strength $s$ increases. This follows the trend of 
the sub-Poissonian statistics, in the sense that the field state becomes more sub-Poissonian as $s$ is increased. 
But differently from what happens to $Q$, there is a different curve for each value of $N$. In fact, the 
parameter $H$ may reach a value $\sim 25 \%$ higher if $N = 3$ compared to when $N = 1$. We may also verify
the dependence of $H$ with the mean photon number of the input state, $\Bar{n}_{th}$, which is shown in Figure
\ref{fig:H_a_n_thermal}. The effect is that the larger the mean photon number of the input state, the closer
the generated state will be to a Gaussian state, as we could expect. We remark that the input state in this
case, the thermal state, is itself a Gaussian state.

\subsection{Phase-diffused coherent state input}

A laser field normally undergoes phase diffusion. If the field state has completely lost its phase information, it can be 
represented by the commonly called phase-diffused coherent state. It is a mixed state diagonal in the Fock basis with a 
Poissonian photon number distribution, or
\begin{equation}
	\rho_n^{(pd)} = e^{-\Bar{n}_{pd}}\frac{\Bar{n}_{pd}^n}{n!},
\end{equation}
where $\Bar{n}_{pd}$ is the mean photon number in the state. The phase-diffused coherent state is not a Gaussian state in the
sense that its Wigner function is not a Gaussian, despite being a positive function. Yet, in this case our quantum scissors device is
also fed by "classical-like" states.

\begin{figure}[htbp]
	\centering
	\includegraphics[scale=0.50]{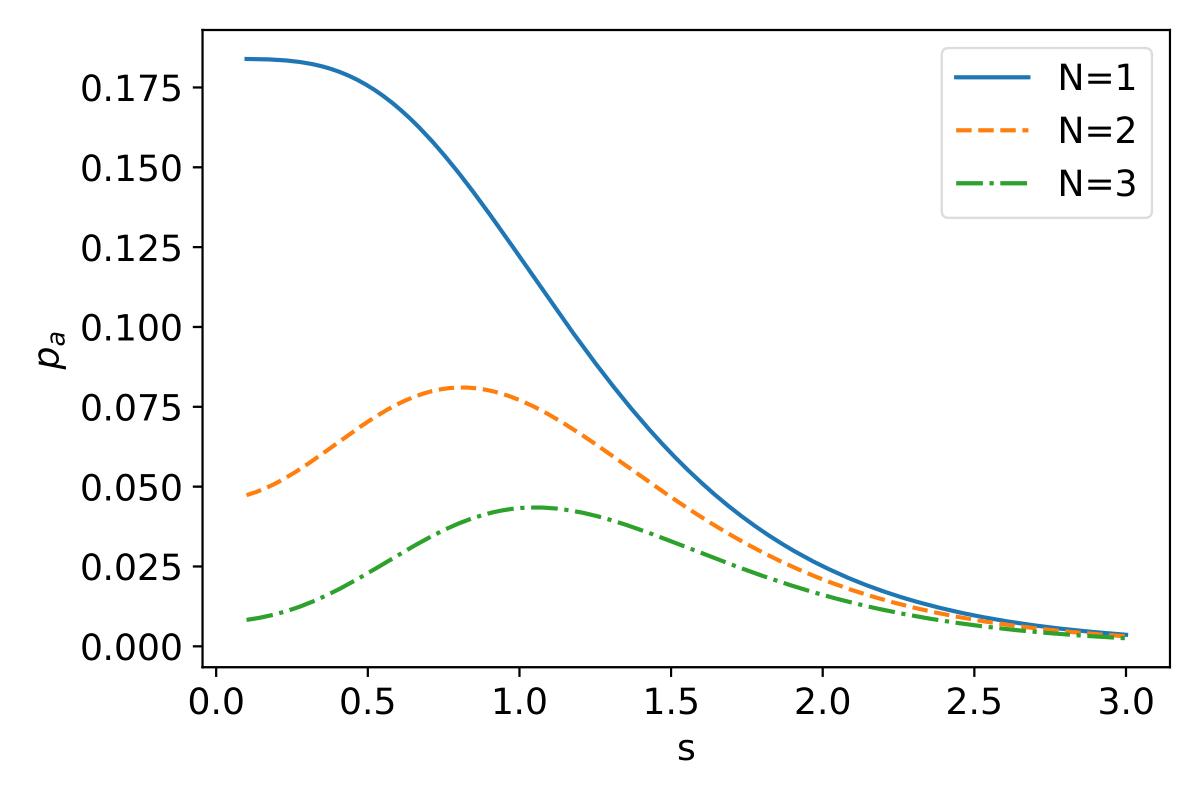}
	\caption{Generation probability $p_A^{(N,0)}$ of states $\hat{\rho}_a^{(N,0)}$ as a function of $s$ for an initial 
		phase-diffused coherent state with $\Bar{n}_{pd} = 1.0$ and a $50 : 50$ beam splitter ($\theta = \pi/4$ rad).}
	\label{fig:prob_pd_a}
\end{figure}

We may calculate the probability of generation of the state $\hat{\rho}_a^{(N,0)}$ as a function of $s$ for an initial 
phase-diffused coherent state with $\Bar{n}_{pd} = 1.0$, a $50 : 50$ beam splitter ($\theta = \pi/4$ rad) and for 
$N = 1,\, 2,\, 3$. This is represented in Figure \ref{fig:prob_pd_a}. The curves have overall shapes similar to the ones 
obtained for the initial thermal state case (see Figure \ref{fig:prob_th_a}).

\subsubsection{Sub-Poissonian statistics: phase-diffused coherent state input}

\begin{figure}[htbp]
	\centering
	\includegraphics[scale=0.50]{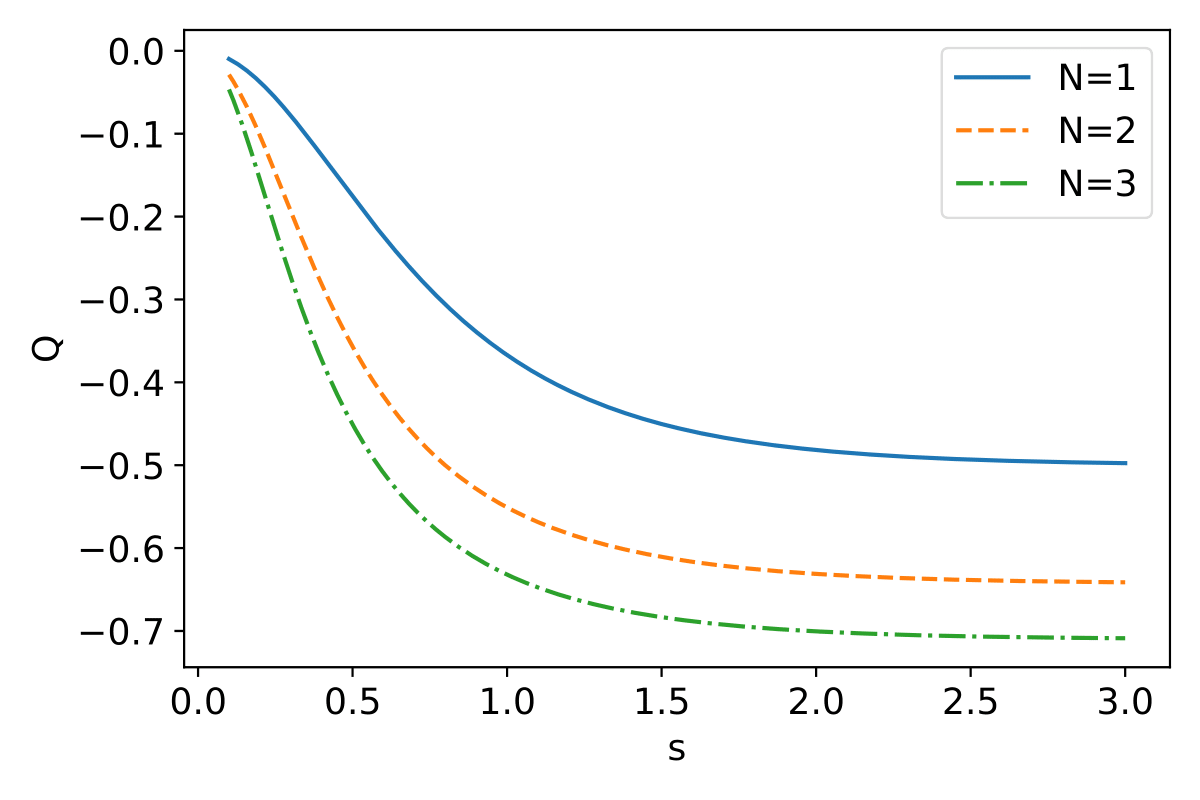}
	\caption{Mandel $Q$ parameter of states $\hat{\rho}_a^{(N,0)}$ as a function of $s$ for an initial 
		phase-diffused coherent state with $\Bar{n}_{pd} = 1.0$ and a $50 : 50$ beam splitter ($\theta = \pi/4$ rad).}
	\label{fig:qparam_a_pdcs}
\end{figure}

\begin{figure}[htbp]
	\centering
	\includegraphics[scale=0.50]{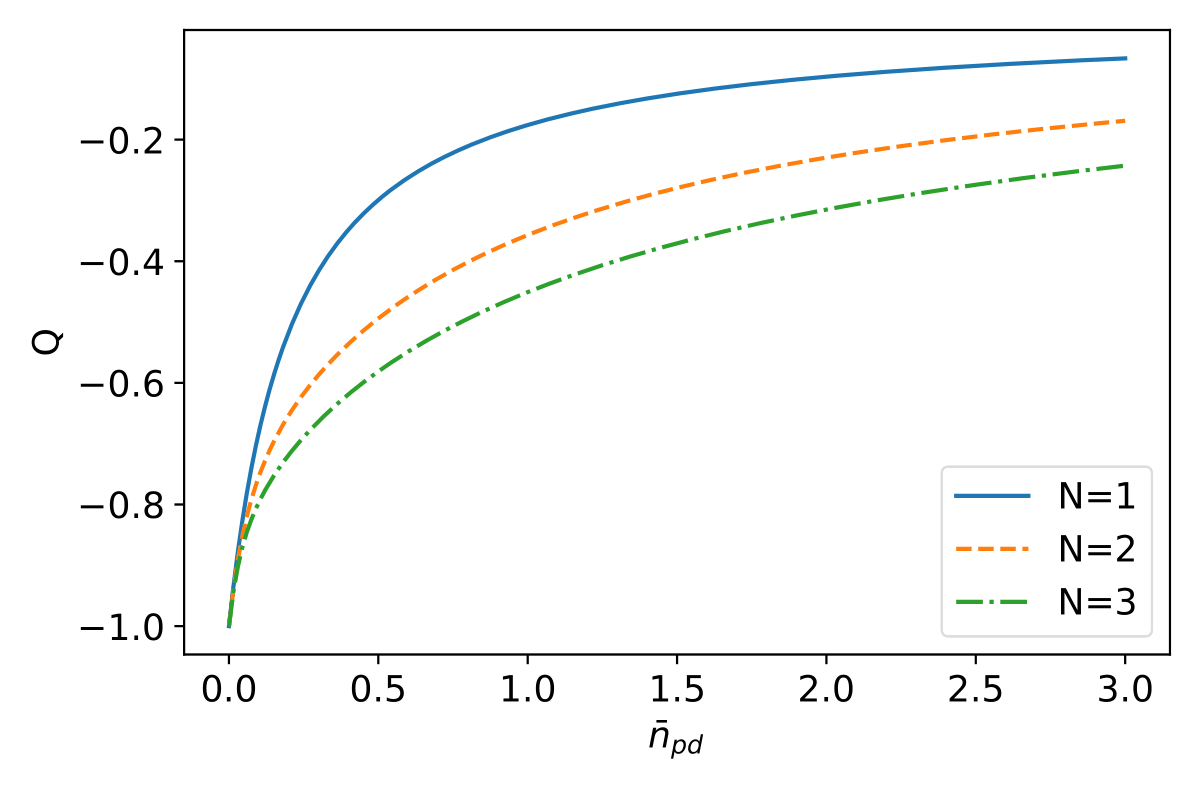}
	\caption{Mandel $Q$ parameter of state $\hat{\rho}_a^{(1,0)}$ as a function of 
		$\Bar{n}_{pd}$ for an initial phase-diffused coherent state, with $s = 0.5$ and 
		a $50 : 50$ beam splitter ($\theta = \pi/4$ rad).}
	\label{fig:qparam_a_n_pdcs}
\end{figure}

The states generated from an initial phase-diffused coherent state are also predominantly sub-Poissonian. 
This is shown in Figure \ref{fig:qparam_a_pdcs}, where we have plotted Mandel's $Q$ parameter relative to the 
state $\hat{\rho}_a^{(N,0)}$ as a function of $s$ for a $50:50$ beam splitter ($\theta = \pi/4$ rad),  
$\Bar{n}_{pd} = 1.0$, and $N = 1,\, 2,\, 3$. In this case, though, differently from the thermal state input example, 
the parameter $Q$ depends on $N$, and states with different degrees of sub-Poissonian character may be obtained
for different values of $N$. For instance, the parameter $Q$ may reach a value $\sim 42 \%$ lower if $N = 3$ 
(more sub-Poissonian) compared to when $N = 1$. We should recall that the success probabilities substantially decrease 
for larger values of the maximum photon number $N$. Interestingly, for an initial thermal state, it is possible
to obtain with the same set of parameters a truncated state more sub-Poissonian than for an initial phase-diffused 
coherent state. This is clearly seen if we compare Figure \ref{fig:qparam_a_s_thermal}
with Figure \ref{fig:qparam_a_pdcs}, e.g., for $N = 1$. Note that in one case we start with a super-Poissonian state 
(thermal), while in the other the initial state is Poissonian (phase-diffused coherent state). Also, the parameter $Q$
plotted as a function of $\Bar{n}_{pd}$ shows a behavior similar to the initial thermal state case, but now 
the sub-Poissonian character has a dependence on $N$, as seen in Figure \ref{fig:qparam_a_n_pdcs}.

\subsubsection{Non-Gaussian character: phase-diffused coherent state input}

\begin{figure}[htbp]
	\centering
	\includegraphics[scale=0.50]{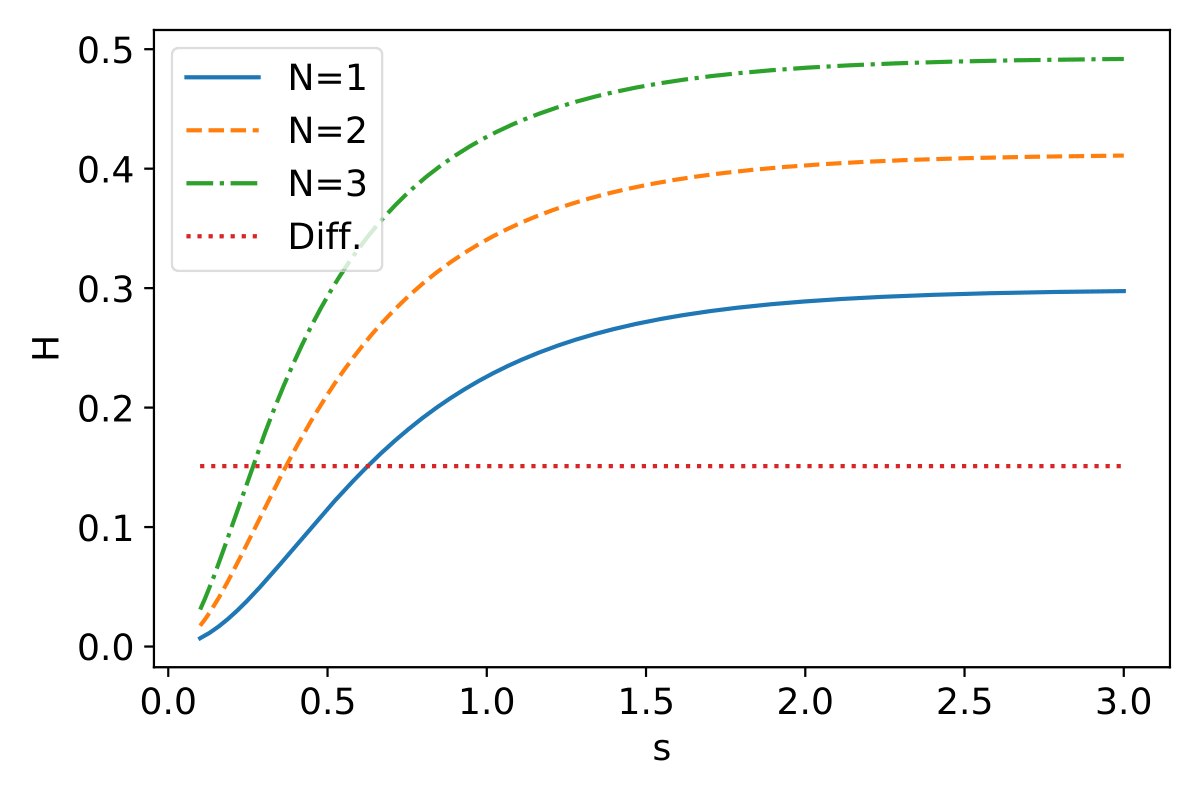}
	\caption{Non-Gaussianity quantifier $H$ of states $\hat{\rho}_a^{(N,0)}$ as a function of $s$ for an initial 
		phase-diffused coherent state with $\Bar{n}_{pd} = 1.0$ and a $50 : 50$ beam splitter ($\theta = \pi/4$ rad).
		The dotted line corresponds to the parameter $H$ of the initial state.}
	\label{fig:H_a_s_pdcs}
	
\end{figure}

\begin{figure}[htbp]
	\centering
	\includegraphics[scale=0.50]{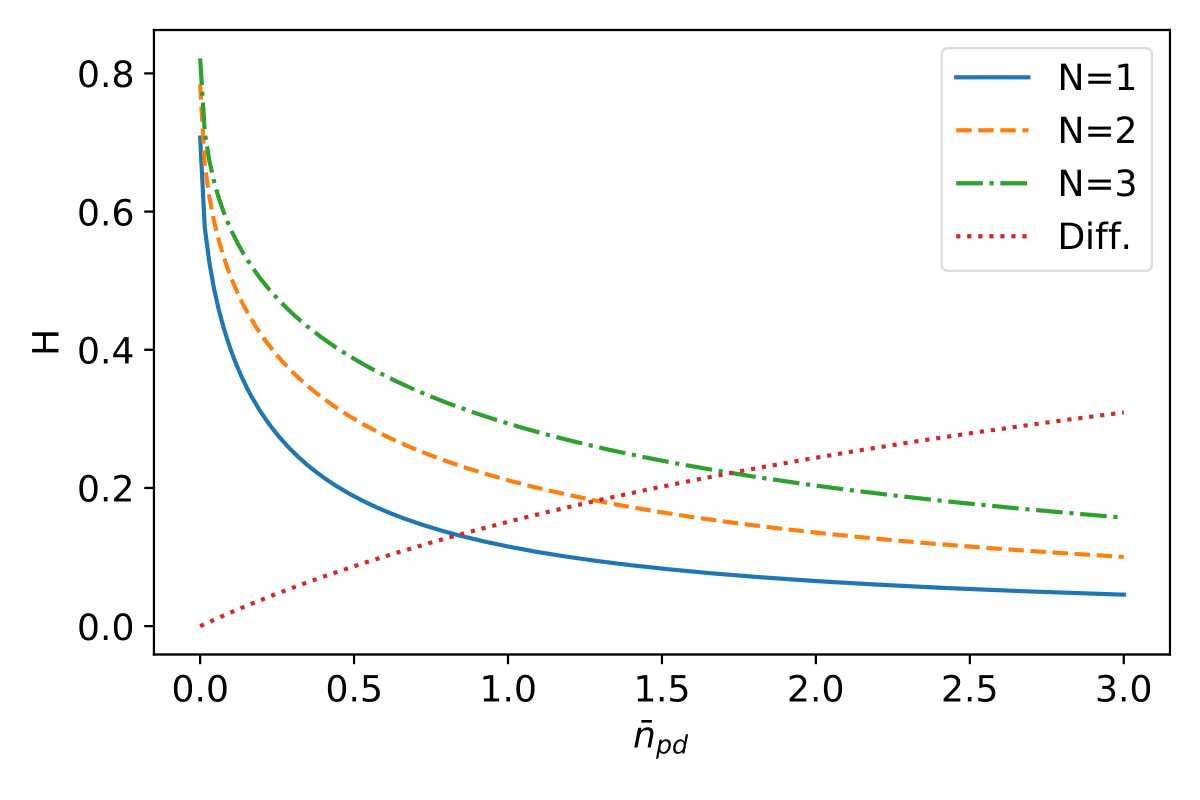}
	\caption{Non-Gaussianity quantifier $H$ of states $\hat{\rho}_a^{(N,0)}$ as a function of 
		$\Bar{n}_{pd}$ for an initial phase-diffused coherent state with $s = 0.5$ and a $50 : 50$ beam 
		splitter ($\theta = \pi/4$ rad). The dotted curve corresponds to the parameter $H$ of the initial state.}
	\label{fig:H_a_n_pdcs}
\end{figure}

The phase-diffused coherent state is not a Gaussian state in the sense that its Wigner function is not
a Gaussian (despite being positive). It is possible to generate states having different degrees
of non-Gaussianity departing from that state as well, as shown next. In Figure \ref{fig:H_a_s_pdcs} we have plotted the 
non-Gaussianity quantifier $H$ of state $\hat{\rho}_a^{(N,0)}$ as a function of $s$ for a $50:50$ beam splitter 
($\theta = \pi/4$ rad) and $\Bar{n}_{pd} = 1.0$. Similarly to thermal state case, the generated field becomes markedly 
more non-Gaussian as the value of the parameter $s$ increases. Also, more intense initial states degrade the non-Gaussian 
character of the generated state, as shown in Figure \ref{fig:H_a_n_pdcs}, where the function $H$ is plotted as a function 
of $\Bar{n}_{pd}$. Again, we note that for an initial thermal state (Gaussian state), it is possible to generate truncated 
states having a parameter $H$ slightly larger, that is, more non-Gaussian than in the case of having an initial phase-diffused 
coherent state. We have also plotted the quantifier $H$ for the initial state, which is itself non-Gaussian, as a function
of $s$ and $\Bar{n}_{pd}$. This is represented by the dotted curves in Figure \ref{fig:H_a_s_pdcs} and Figure \ref{fig:H_a_n_pdcs}, 
respectively.

\section{Conclusions}
\label{section4}

In this work, we have studied the generation of nonclassical states from mixed input states using a modified 
quantum scissors device, consisting of a nondegenerate parametric amplifier, a beam splitter and two photodetectors. 
In this scheme, the input state is converted into a truncated state in the Fock basis. Two types of state can be
generated: i) states having a maximum Fock number $N$, or ii) states having a minimum Fock number $N$, depending
on the placement of the photodetectors in relation to the output ports. One advantage of our scheme is that requires
no more than simple input states such as vacuum states and a classical pump to operate. In this setup, the 
interaction within the parametric amplifier provides the necessary nonclassical resource, as one of the modes of
the generated two-mode squeezed vacuum state enters the input port $\hat{b}$ of the beam splitter. 
Regarding the field state to be truncated, we may not have at our disposal an ideal source of pure states
to generate the inputs, but rather, noisier (mixed) states. Here, we investigated the consequences of having input mixed states
diagonal in the Fock basis, e.g., the thermal states and the phase-diffused coherent states. We demonstrated that truncated (mixed) 
states having sub-Poissonian statistics as well as non-Gaussian character can be generated with non-negligible success probabilities. 
The nonclassical properties of the generated states may be readily controlled via the parametric amplifier strength $s$, as well 
as the beam splitter's transmittance $\mathtt{T}$. However because of the mixed state input, the phase $\phi$ of the pump will play 
no role. Thus, unlike what happens in the case of pure input states \cite{scissors21}, this is not a phase-sensitive process, and 
there will be no quadrature squeezing at all. Interestingly, we found that it is possible to obtain states having more pronounced 
nonclassical features from the thermal states, which are Gaussian and super-Poissonian, compared to phase-diffused coherent states 
(especially for $N = 1$). We have so far discussed here the ideal situation in which are employed perfect photodetectors. 
Nevertheless, based on the analysis already made in \cite{scissors21}, we can expect that the effects of a non-unitary quantum 
efficiency of the detectors and dark counts will be similar also for mixed input states, i.e., it should be possible to
generate truncated states with relatively high fidelity in this case too, even in a non-ideal situation.

\section*{Acknowledgments}
E.P.M. would like to acknowledge financial support from CAPES (Coordenadoria de Aperfeiçoamento
de Pessoal de Nível Superior), Brazil, grant N${\textsuperscript{\underline{o}}}$ 88887.514500/2020-00.
This work was also supported by CNPq (Conselho Nacional de
Desenvolvimento Científico e Tecnológico, Brazil), via the INCT-IQ
(National Institute for Science and Technology of Quantum Information),
grant N${\textsuperscript{\underline{o}}}$ 465469/2014-0.

\bibliographystyle{unsrt}
\bibliography{refsscissors}

\newpage


\end{document}